\documentclass{revtex4}
\usepackage{psfig}
\usepackage{subfigure}
\usepackage{graphicx}
\usepackage{doublespace}

\begin{document}

\renewcommand{\baselinestretch}{2}


\title{Effects of the network structural properties on its controllability}
\author{Francesco Sorrentino${}^{*\ddagger}$}
\affiliation{${}^*$University of Naples Federico II, Naples 80125,
Italy \\ and Institute for Research in Electronics and Applied Physics, University of Maryland, College Park, Maryland 20740, USA\\ ${}^\ddagger$ E-mail: {\tt fsorrent@unina.it}}
\begin{abstract}
In a recent paper, it has been suggested that the controllability of
a diffusively coupled complex network, subject to localized feedback loops at
some of its vertices, can be 
assessed by means of a
Master Stability Function approach, where the network controllability is defined in terms of 
the spectral properties of an appropriate Laplacian
matrix. 
Following that approach, a comparison study is reported here among
different network topologies in terms of their controllability. The
effects of heterogeneity in the degree distribution, as well as
of degree correlation and community structure, are discussed.

\end{abstract}
\maketitle

\textbf{In recent years, synchronization of complex networks of
coupled dynamical systems has become a research subject of increasing attention within
the scientific community. This is partly motivated by the frequent
observation of synchronization phenomena in a wide range of different
contexts, ranging from biology to medical and social sciences. On
the other hand, this is because synchronization is considered a paradigmatic example of phase transitions, 
that
under certain circumstances may occur when ensembles of dynamical
systems are coupled together.}

\textbf{ Generally speaking, several different processes may
lead to the onset of synchronization.  
Namely, it may be either the result of a self organized process occurring when several coupled dynamical systems, starting from
different initial conditions, converge in the same dynamical
evolution (which is in general not known a priori); or it can be
provoked by some feedback loops driving the set of all the
dynamical systems toward a desired predetermined reference
evolution.  In this paper, we focus on synchronization of diffusively coupled networks. Moreover, we consider that a subset of nodes are selected to be controlled. 
This assumption is motivated by real-world networks
observation, where a decentralized control action is often applied
only to part of the nodes. For instance, pacemaker
cells 
have been observed to regulate several functions in living organisms; other examples are present in human networks, where particular individuals, called \emph{leaders}, have been observed to be capable
to influence the network collective dynamics. }

\textbf{  
Here, the controllability of a given
complex network, i.e., its propensity to being controlled onto a
given reference evolution by means of a decentralized control action, is defined as the width of the range of the coupling strength term among the oscillators, which  stabilizes the reference evolution. 
A detailed comparison  among different complex
networks in terms of their controllability,  characterized by different degree distributions, degree
correlation properties, as well as community structure, will be
reported.} 

\section{Introduction}

The control of the complex dynamics
which take place on networks of many interconnected units is an issue
of primary importance in various fields of applied sciences.
In a recent paper \cite{PC}, the problem of
how a dynamical complex network of diffusively coupled systems, can
be controlled onto a synchronous evolution, was studied by applying a local
feedback action to a small portion of the network nodes.

In nature, there are many situations where the control of a very large
complex network is an important functional
requirement; this is the case, e.g., of some bodily
functions, such as the contemporaneous beats of the heart cells
\cite{PesBOOK}, or the synchronous behaviors of the cells of the
suprachiasmatic nucleus in
the brain, 
which sets the clock of the circadian bodily rhythms \cite{supra}.
Other examples can be easily found in social networks, where the
formation of mass-opinions and the emergence of collective behaviors
are frequently observed. Generally speaking, this issue is
particularly relevant to those situations where a given common
behavior of all the network microscopic systems represents a
functional requirement for the network dynamics at the macroscopic
level.

It is worth noting here that sometimes, in the literature, the same phenomenon is also
referred to as the \emph{entrainment} of a network
of dynamical systems; however in what follows, for simplicity, the common term control is used.

In \cite{PC}, the problem of network controllability was studied
via a Master Stability Function approach \cite{Pe:Ca}.
Under the hypothesis of all the network dynamical systems being
identical, and the coupling being diffusive, a quantity was defined
to assess the propensity of any given complex network (or lattice)
to being controlled. In so doing, the network controllability was
defined as a structural property, independent of the particular type
of dynamics considered at the network nodes. Specifically, the
network controllability was measured in terms of a simple matrix
spectral index (to be precisely defined below).

This issue seems closely related to that of the network
synchronizability, as studied e.g. in
\cite{Pe:Ca,Ni:Mo,Bocc2,report}; indeed, in both cases the object of
study is the range of values of a defined control parameter that
rules the stability of the synchronous/reference evolution. However, there is a fundamental difference. The network
synchronization problem considers systems which autonomously settle
onto a generic (not assigned a priori) synchronous evolution. In
such a case, synchronization is achieved by means of a
self-organized process, where the involved dynamical systems
dynamically adjust their trajectories until they eventually converge
onto the same evolution. Many phenomena in nature resemble this kind
of behavior, for example, the spontaneous emergence of synchronous
behavior observed in populations of fireflies \cite{SYNCBOOK}.

On the other hand, a different phenomenon (kind of dynamics) takes
place when an entity external to the network, or some
of its nodes (which do not undergo the influence of the other
network systems) are deputed to control the whole network onto a
desired synchronous evolution. Specifically, in this case, part of
the network dynamical systems 
drive all the other network systems toward their own dynamical
evolution. Moreover this difference is not only phenomenological. In
mathematical terms, this corresponds to the set of the solutions belonging to the synchronization manifold $\{x_1=x_2=...=x_N\}$,
shrinking to the only admissible solution $\{x_1=x_2=...=x_N=s(t)\}$, where $s(t)$ is the
reference evolution
chosen for the network. 
Furthermore there is another difference. In fact, the number of
network dynamical modes that need to be ensured to be stable, varies
when assessing the synchronizability/controllability of a given
complex network (for more details, see
\cite{PC} and Sec. II).

Examples of control are the synchronous beats of the heart cells
that are 
regulated by the activity of the pacemaker cells situated at the
sinoatrial node \cite{PesBOOK} 
and the circadian rhythms
, observed in many living organisms, 
entrained by the light-dark cycle (for humans it has been shown that
the intrinsic period of oscillations of the cells in the
suprachiasmatic nucleus is different from the 24-hour cycle). 

Sometimes, when looking at real-world phenomena, the distinction
between these two different behaviors is not so obvious. For
instance, in \cite{McC71} the phenomenon of menstrual
synchronization among female roommates and close friends was
reported and it was suggested that this could be understood in terms
of mutual pheromonal  
interactions among individuals living together or interacting closely. In a
successive publication \cite{Rus80}, it was reported the case of a
female subject, whose cycle was
very regular, being 
able to lock other women cycles
on hers (for a discussion see also \cite{SYNCBOOK}.)

Recently, several papers in the physics literature have dealt with 
the issue of controlling complex networks. In \cite{Mikh1,Mikh2},
for example, a network of identical Kuramoto oscillators
is entrained by a single pacemaker, characterized by a different
frequency from that of the other oscillators (for a discussions on
the effects of the network topology on the synchronization of
networks of Kuramoto oscillators, see \cite{restr06,restr05}).


Because of the distributed nature of complex networks, whose dynamics
are mainly decentralized, 
it is feasible to control them by acting
locally 
on part of their nodes 
and exploiting the coupling effects between these and the rest of the
network to achieve the desired goal. In pinning control schemes
\cite{Pinn1,Pinn2,PinnA,PinnIEEE}, some nodes are permanently
selected (pinned) to be the network controllers.  Specifically,
these nodes, referred to as reference sites or pinned sites, play the
role of network \emph{leaders}/\emph{pacemakers}.

The equations for a diffusively coupled complex network under the effect of pinning control, can be generally formulated as
follows:
\begin{equation}
\frac{dx_i}{dt}=f(x_i)+\sigma \sum_{j=1}^{N} {\mathcal L}_{ij}
h(x_j)+ \sigma \kappa_i B_i (s-x_i), \label{eq:net2}
\end{equation}
$i=1,...,N$, representing the behavior of $N$ identical dynamical
systems coupled through the network edges.

The first term on the right hand side of (\ref{eq:net2}) describes the
state dynamics of the oscillator at each node, $\{ x_i(t),
i=1,...,N\}$, via the nonlinear vector field $f(x_i)$; the second
term represents the coupling among pairs of connected oscillators,
through a generic output function $h(x_i)$, where the coupling gain
$\sigma$ represents the overall strength of the interaction.
Information about the weighed network topology is contained in the
Laplacian matrix $\mathcal L$, whose entries $\mathcal{L}_{ij}$, are
zero if node $i$ is not connected to node $j \neq i$,  but are
negative if there is a direct influence from node $i$ to node $j$,
with $\mid \mathcal{L}_{ij} \mid$ giving a measure of the strength
of the interaction, and $\mathcal{L}_{ii} =-\sum_j \mathcal{L}_{ij}$,
$i=1,2,...,N$, representing the diffusive coupling. In what follows, assume that the network is globally
connected, which ensures the matrix $\mathcal{L}$ have only one zero eigenvalue. 
The control action is directly applied only to the reference nodes,
which are indexed by the entries of the binary vector $B$: $B_i=1$
($B_i=0$) if node $i$ is controlled (not controlled). Hereafter,
we assume $\sum_i B_i\geq 1$. As commonly assumed in  pinning
control schemes, such nodes play the role of leading the others
toward the desired reference evolution, say $s(t)$.

Here the control input is generated by a simple
state-feedback law with respect to the reference evolution $s(t)$,
which is assumed to satisfy $\frac{ds}{dt}=f(s)$, and $\kappa_i$ is
the control gain acting on node $i$. Note that, even though there is
no reason for considering that the control gains cannot vary among the
reference sites, for the sake of simplicity in the rest of the
paper, we set $\kappa_i=\kappa$ $\forall i$ , 
i.e., we assume they are the same at all the reference sites.

Another problem of non-negligible importance is represented by the
choice of the nodes to be pinned from the set of all the network vertices. First of all, one should decide the number of nodes
$m=\sum_i B_i$ to control; in what follows it is assumed that the
number of controlled nodes is ruled by the pinning probability
$p=m/N$. In many real situations, this decision is often affected
by some environmental constraints. In particular, when dealing with
biological networks, it becomes particularly evident that both the
controlled and the uncontrolled nodes play different but evenly
important functions. For example, in the heart, pacemaker and
non-pacemaker cells exhibit different phases and amplitudes 
of their pulsations.

Note that once the number of pinned nodes $m$ is given, there are
$  {N} \choose {m}$ different
                              possibilities of choosing the nodes to
                              control.
Usually, the two following strategies
for choosing the pinned nodes are considered: (i) \emph{Random
pinning}: The $m$ pinned nodes are randomly selected with uniform
probability from the set of all the nodes. (ii)
\emph{Selective pinning}: The $m$ pinned nodes are first sorted
according to a certain property of the nodes, for
instance, the nodes degree or betweenness centrality, then the
pinned nodes are chosen in that particular order.

The rest of the paper is outlined as follows. In Sec. II, a
definition of controllability is presented for a general complex dynamical networks, subject
to a decentralized control action. In Sec. III, the
effects of heterogeneity in the network degree distribution are
discussed. The role of degree correlation is further analyzed in
Sec. IV. In Sec. V, linear and square lattices are considered, showing that the distance among the selected reference nodes
across the network is an important property to the network
controllability. This idea is confirmed by simulations of complex
networks with community structure in Sec. VI.

\section{A structural measure of network controllability}

Following \cite{PC}, here we are interested in the stability of the
solutions $x_1(t)=x_2(t)=...=x_N(t)=s(t)$ of the network (1). After standard
manipulations \cite{Pe:Ca,PC}, this can be evaluated in terms of the
dynamics of $N$ independent blocks in the parameters $\alpha=\sigma
\mu_i$, $i=1,...,N$ \cite{Pe:Ca}:

\begin{equation}
\frac{d \eta_i}{dt}= [Jf(s)-\alpha Jh(s)] \eta_i, \quad i=2,...,N
\label{blocks},
\end{equation}
where $Jf(s)$ and $Jh(s)$ are the Jacobians of the functions $f$ and
$h$ calculated about the time varying reference evolution $s(t)$ and
$\mu_i$, $i=1,...,N$ are the eigenvalues of the $N$-dimensional
structural matrix
\begin{eqnarray}\mathcal{M}= \{\mathcal{M}_ {ij} \}=
{\small\small\small{ \pmatrix{ \mathcal{L}_{11}+ B_1 \kappa_1 &
\mathcal{L}_{12} & \cdots & \mathcal{L}_{1N}  \cr \mathcal{L}_{21} &
\mathcal{L}_{22}+B_2 \kappa_2 & \cdots & \mathcal{L}_{2N}  \cr
&\ddots & & \vdots \cr \mathcal{L}_{N1} & \mathcal{L}_{N2} & \cdots
& \mathcal{L}_{NN}+ B_N \kappa_N \cr }}} \nonumber
\end{eqnarray}.

Hereafter, assume the network to be undirected (and unweighed),
which ensures the matrix $\mathcal{M}$ be symmetric and thus its
spectrum be real \footnote{We wish to emphasize that the
assumption of undirected and unweighed networks, is done here for
the sake of simplicity and without loss of generality. The MSF
approach, presented in this paper, is also valid under very general
assumptions about the network topology (by allowing eventually the
spectrum of the matrix $\mathcal{M}$ to be complex \cite{Bocc2} and
even the matrix $\mathcal{M}$ to be non-diagonalizable
\cite{Ni:Mo06})}. Suppose its eigenvalues are sorted as $\mu_1
\leq \mu_2 \leq .... \leq \mu_N$. Note that $\mathcal{M}$ is not a
Laplacian matrix; however, as explained in \cite{PC}, the same
analysis can be performed in terms of an $(N+1)$-dimensional Laplacian
matrix, having the same spectrum as $\mathcal{M}$, plus one additional zero
eigenvalue (for a comparison, the reader is referred to \cite{PC}).

Specifically, when all the Lyapunov exponents $\Lambda(\alpha)$
associated with the $N$ systems in (\ref{blocks}) are negative, the
trajectories of all the network systems are found to be stable
about the reference evolution $s(t)$. Note that differing from
the previously reported case of the network synchronizability, this
requires the investigation of one more eigenvalue, the one
associated with the dynamics along the direction of the
synchronization manifold. 
It is worth noting that theoretically,
it is this additional eigenvalue that makes the difference between
the synchronizability and the controllability of complex dynamical
networks. Moreover, different from the case of the network synchronization, in order to control a network, it is not necessary to have a unique globally connected cluster, while this condition can be replaced by the other one, that in each cluster at least one controller is present (for more details, see Sec. VI).

Now introduce the Mater Stability Function (MSF) that 
associates to each value of the normalized coupling strength
$\alpha$, the largest Lyapunov exponent $\Lambda(\alpha)$ of any
given system of the form (\ref{blocks}). Note that the particular
MSF depends only on the choice of the dynamical functions $f$ and
$h$ but not on the network topology. Moreover, once the MSF is
assigned, it is 
possible to define an interval of values
of $\alpha$, say $[\alpha_{min}, \alpha_{max}]$, which corresponds to
negative values of the MSF.

In what follows, we will distinguish between the two following different situations: (i) when $\alpha_{max}$ is finite, the width
of the range of values of $\sigma$, say $\Sigma$, corresponding to
a negative MSF, is finite and is an increasing function of the
eigenratio $R=\mu_N/\mu_1$; (ii) when
$\alpha_{max}$ is infinite, 
the sole
$\mu_1$ gives information about the network controllability: namely,
the higher $\mu_1$ is, the more the network is controllable, i.e., the
lower is the critical value of $\sigma$ above which the reference
evolution is stable (note that this fits perfectly the
global stability conditions obtained in \cite{PinnIEEE}).


The most important result of this approach, is the decoupling of the
structural information about the network topology from some
particular kind of dynamics at the network nodes. Here, the network structural properties are not only
the network topology in terms of the connections and
the weights over them, but also the particular choice of the
reference sites and the control gains over them (i.e., all the
information encoded in the matrix $\mathcal{M}$). This indicates
that the network controllability, different from the network
synchronizability, can be enhanced by an appropriate choice of the
reference sites and of the control gains over them. Therefore, in
the rest of the paper, the design of 
the most effective strategies will be discussed as how to place the controllers over a
given network to enhance its controllability.

As a by-product, it becomes possible to compare directly the
controllability of different network topologies in terms of the
spectral properties of the matrix $\mathcal{M}$.  Note that in so
doing, one needs to
distinguish the two 
cases (i) and (ii), as indicated above. Specifically, in what
follows, the network dynamical systems will be classified respectively as class I or class II 
according to these two cases.

As explained in \cite{PC}, by following this approach it is
even possible to compare the controllability and the
synchronizability of a given network. In particular, it is possible
to compare the widths of the intervals of the coupling gain $\sigma$
that lead to stability of the synchronous (or reference)
evolutions. However, as shown in \cite{PC}, in the case that the
number of reference sites $m \ll N$, the
networks are generally harder to control than to synchronize.

A surprising finding in \cite{PC} is that for class I systems, the network controllability is reduced, as the average control gain
$\kappa$ is increased to above a certain value, and this property was
found over a wide variety of different networks and
even lattices.

\section{Effects of heterogeneity in the degree distribution}

Heterogeneity in the degree distribution is probably the most
important feature that characterizes the structures of real networks.
The discovery that the basic structure of many real-world networks is
characterized by a power-law degree distribution, was pointed out by Barabasi and Albert in their seminal work
\cite{Ba:Al99}, which has been verified by many observations of real networks. Specifically, the
analysis of data sets of biological, social and technological
networks has showed that these typically exhibit a
power-law degree distributions, $P(k) \sim k^{-\gamma}$, which is
characterized
by high heterogeneity. 

In \cite{Ni:Mo}, the Master Stability Function method was used to
assess the synchronizability of networks characterized by different
degree distributions, and a surprising observation was that the
higher the network heterogeneity, the lower their synchronizability
. This indicates that the range of values of the coupling strength
$\sigma$, for which real networks can be synchronized, is
particularly narrow with respect to other network architectures. In \cite{Ni:Mo}, this interesting phenomenon was called the
\emph{paradox of heterogeneity}.

In this section  we attempt to assess the controllability of a
complex network by varying the heterogeneity of its degree
distribution. In order to reproduce various networks characterized by
different degree distributions, we introduce an appropriate network
construction model.

Specifically, we generate a scale-free random network through the
static model described in what follows. Firstly, startup with a network of size $N$, assign to each vertex
$i=1,2,...,N$, a weight $w_i=(i+\theta)^{-\mu}$, where the so-called Zipf
exponent $\mu$ lies in the range $[0,1)$ and $\theta \ll N$.  
Assume that initially no edges are present among the network
vertices, then edges are added one by one until $\mathcal{E}$ connections are created. For each new edge,
two vertices are randomly selected, each one with probability proportional to its weight, and
they are connected unless a link already exists or the two
selected nodes are the same.

By following this construction, the expected degree of node $i$, say
$k_i$, is simply equal to
$k_i=c (i+\theta)^{-\mu}$, where $c=2\mathcal{E}/\sum_{j=1}^{N} (j+\theta)^{-\mu}$. Thus 
\begin{equation}
P(k_i \leq \overline{k})=P(c (i+\theta)^{-\mu} \leq \overline{k})=
P\Big(i+\theta>
\Big( \frac{\overline{k}}{c} \Big)^{-\frac{1}{\mu}}\Big)=\frac{N+\theta-(\frac{\overline{k}}{c})^{-\frac{1}{\mu}}}{N-1}.
\end{equation}

Then the probability of finding a vertex of degree $\overline{k}$ is
$P(\overline{k})= c' \overline{k}^{-(1+\frac{1}{\mu})}$, with
$c'=c^{\frac{1}{\mu}}/(\mu (N-1))$ and hence the power-law scaling is
satisfied with $\gamma= 1+ \frac{1}{\mu}$. Note that the scaling
of the degree distribution is independent of $\theta$.

The methodology presented above is an extension of the classical
\textit{static model} (which has been intensively studied in a number of
papers), introduced in \cite{korea}; namely this is recovered in the
particular case of $\theta=0$. 

\begin{figure}[t]
\centerline{
\psfig{figure=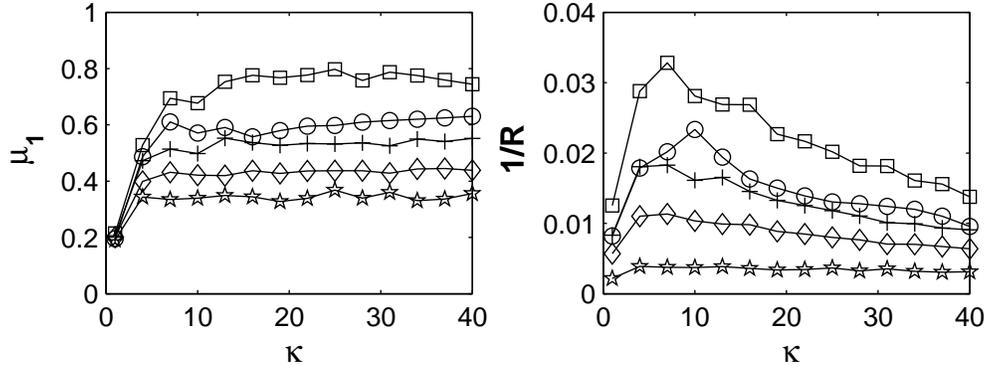,width=15cm}}
\caption{ \small Networks of $N=10^3$ nodes, $\mathcal{E}=3 \times
10^3$ edges, $\theta=10$. $\mu_1$ and $1/R$ are plotted versus
$\kappa$ in networks characterized by different degree distributions of the type
$P(k) \sim k^{-\gamma}$. The legend is as follows: $p=0.05,
\gamma=2.1$ (stars); $p=0.05, \gamma=3$ (diamonds); $p=0.05,
\gamma=4$ (plus); $p=0.25, \gamma=2.1$ (circles); $p=0.25, \gamma=3$
(squares). \label{paradox1}}
\end{figure}

 The main results are shown in Fig. \ref{paradox1}, where $\mu_1$ and
$1/R$ have been plotted versus the coupling gain $\kappa$ for
networks characterized by different power-law exponents \footnote{
In particular, in the case where the network resulted not globally
connected (i.e. some isolated nodes or clusters of few nodes
connected together appeared), the spectral properties of the giant
component have been considered.} (i.e. $\gamma=2.1,3,4$ ) and in the
cases of a small/large number of controlled nodes (in terms of the
probability $p=0.05,0.25$). Remind that the larger $1/R$ and
$\mu_1$ are, the more the networks of dynamical systems in class I and
class II, are controllable.

Fig. \ref{paradox1} clearly shows that by increasing $\gamma$, it is
possible to enhance the network controllability. This result is
particularly surprising, since diffusion dynamics are known to be
favored in networks characterized by higher heterogeneity in the
degree distribution \cite{Pa:Ve00a,restr06}. Moreover this is in
accordance with the phenomenon known as the paradox of
heterogeneity in the context of network
synchronization \cite{Ni:Mo}. Thus 
the paradox of heterogeneity affects not only the synchronizability
of complex networks, but also their controllability.

Dynamical simulations were carried out involving R\"ossler
oscillators diffusively coupled in the $x-z$ variables (which are known
to belong to class I; see also \cite{PC}). Namely we have
considered $N$ identical R\"ossler oscillators placed at the network
vertices; the dynamics at each node $i$ is described by the
following vector field: $f(x_i)=f(x_{i1},x_{i2},x_{i3})=(- x_{i2}\;
-\; x_{i3}, \quad x_{i1}\;+\;0.165 x_{i2}, \quad
0.2+(x_{i1}-10)x_{i3})$. The output function $h$ has been chosen, as
in \cite{PC}, to be $h(x)=Hx$, where $H$ is the matrix, {\small
$\pmatrix{ (1 & 0 & 0) , (0 & 0 & 0),  (0 & 0 & 1) }$}, indicating
that the oscillators are coupled through the variables $x_{i1}$ and
$x_{i3}$  ($i=1,2,...,N$). The asymptotic value of the control
error $E=\frac{1}{(\Delta T) N} \sum_{i=1}^{N} \int_T^{T+\Delta T}
\| x_i(t)-s(t)\| dt$, with $\| x \|= |x_{1}|+|x_{2}|+|x_{3}|$, has
been computed under variations of the control gain $\kappa$ and the
number of controlled nodes $m$.

\begin{figure}[t]
\begin{center}
\centerline{
\psfig{figure=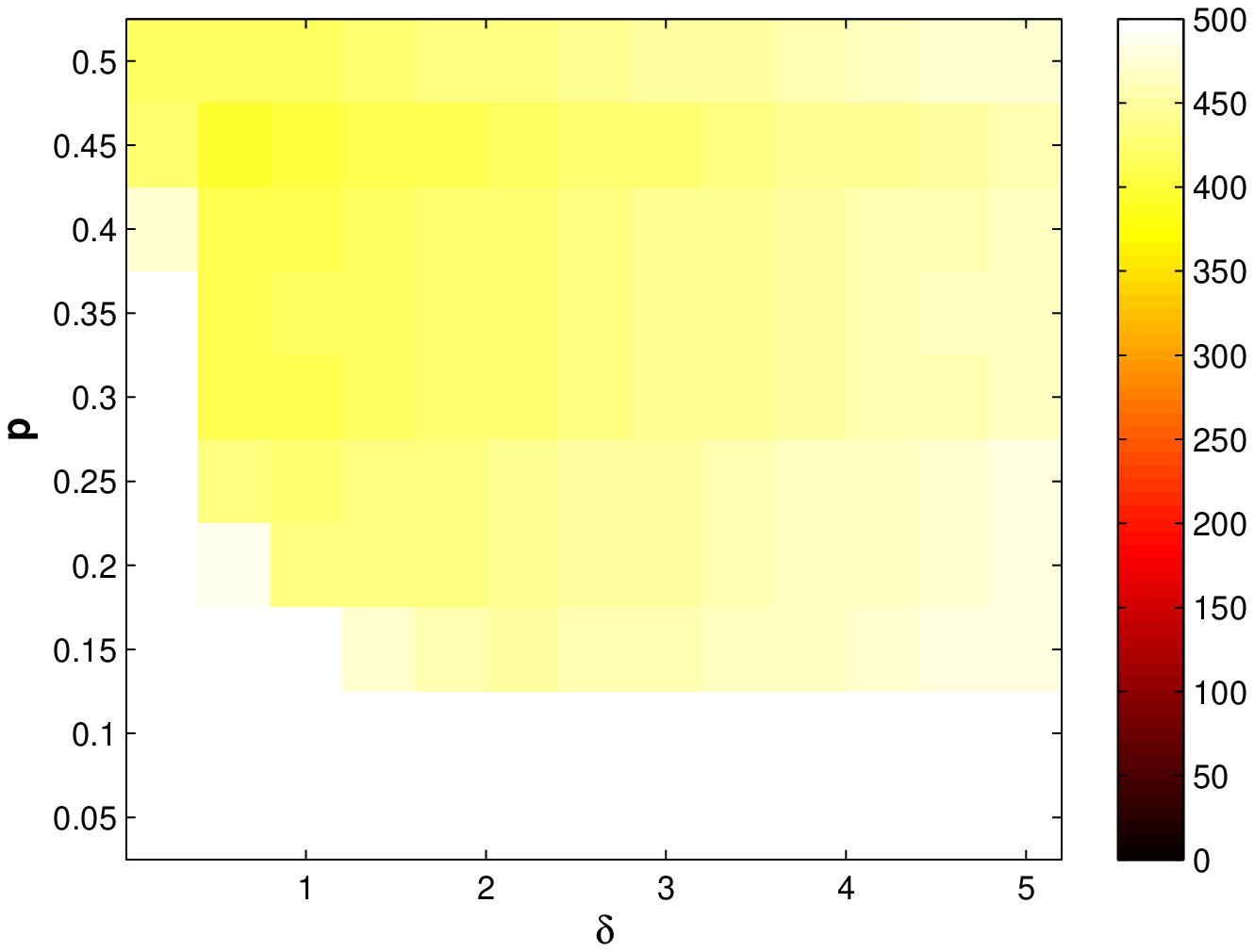,angle=0,width=9cm}
\psfig{figure=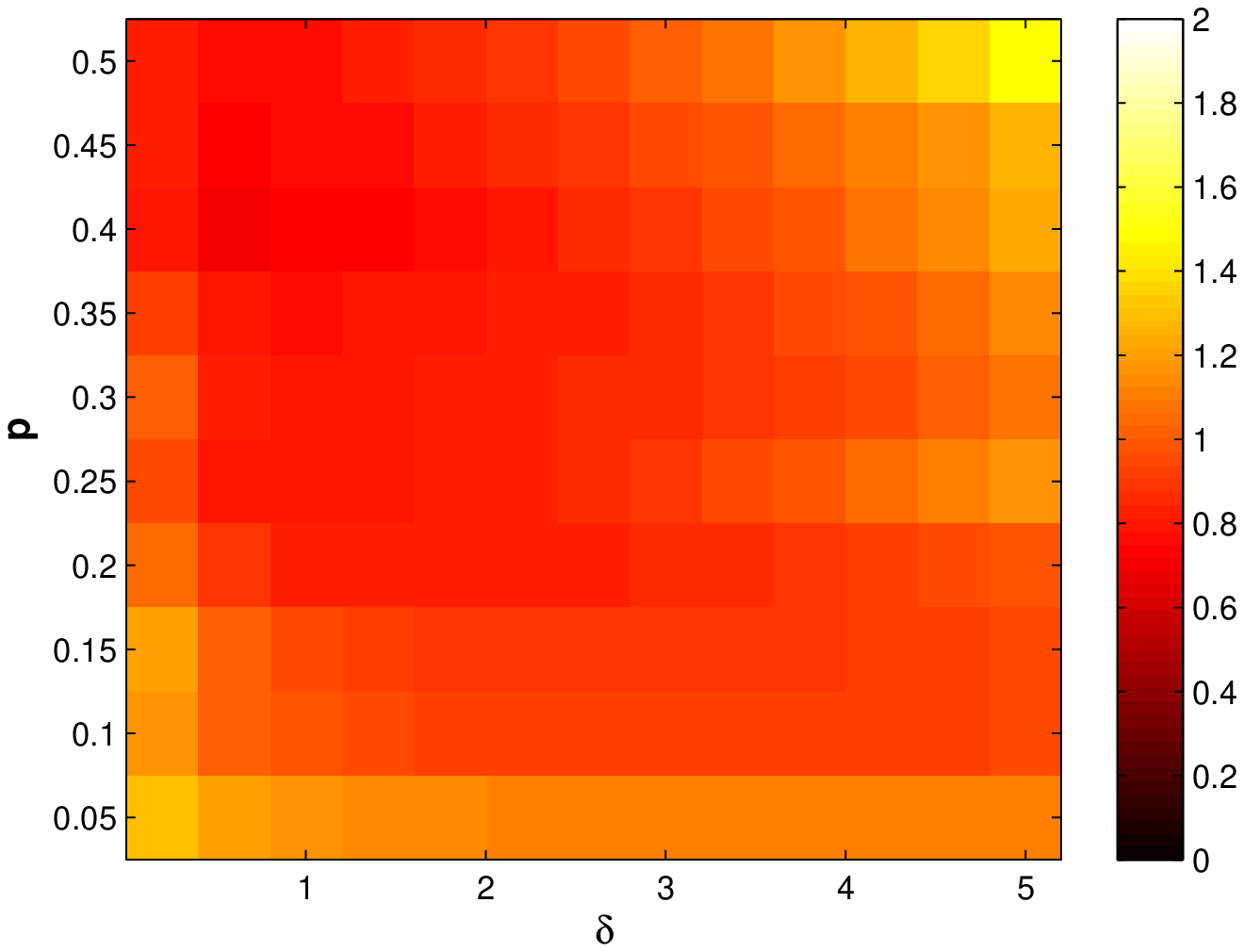,angle=0,width=9cm}}
\caption{\label{MAIN} \small A scale free network of $10^3$ nodes with degree
distribution exponent $\gamma=2.1$ is considered. The left figure shows
the eigenratio $R$ as varying both the control gain $\delta=\sigma
\kappa$ ($\sigma=0.2$) and the number of controlled nodes (in terms
of the probability $p$).  The right picture shows the control
error at regime $E$ as function of the control gain $\delta= \sigma
\kappa$ and the probability $p$ under the same conditions as in the
left plot. \label{paradoxD12}}
\end{center}
\end{figure}

\begin{figure}[t]
\centerline{\psfig{figure=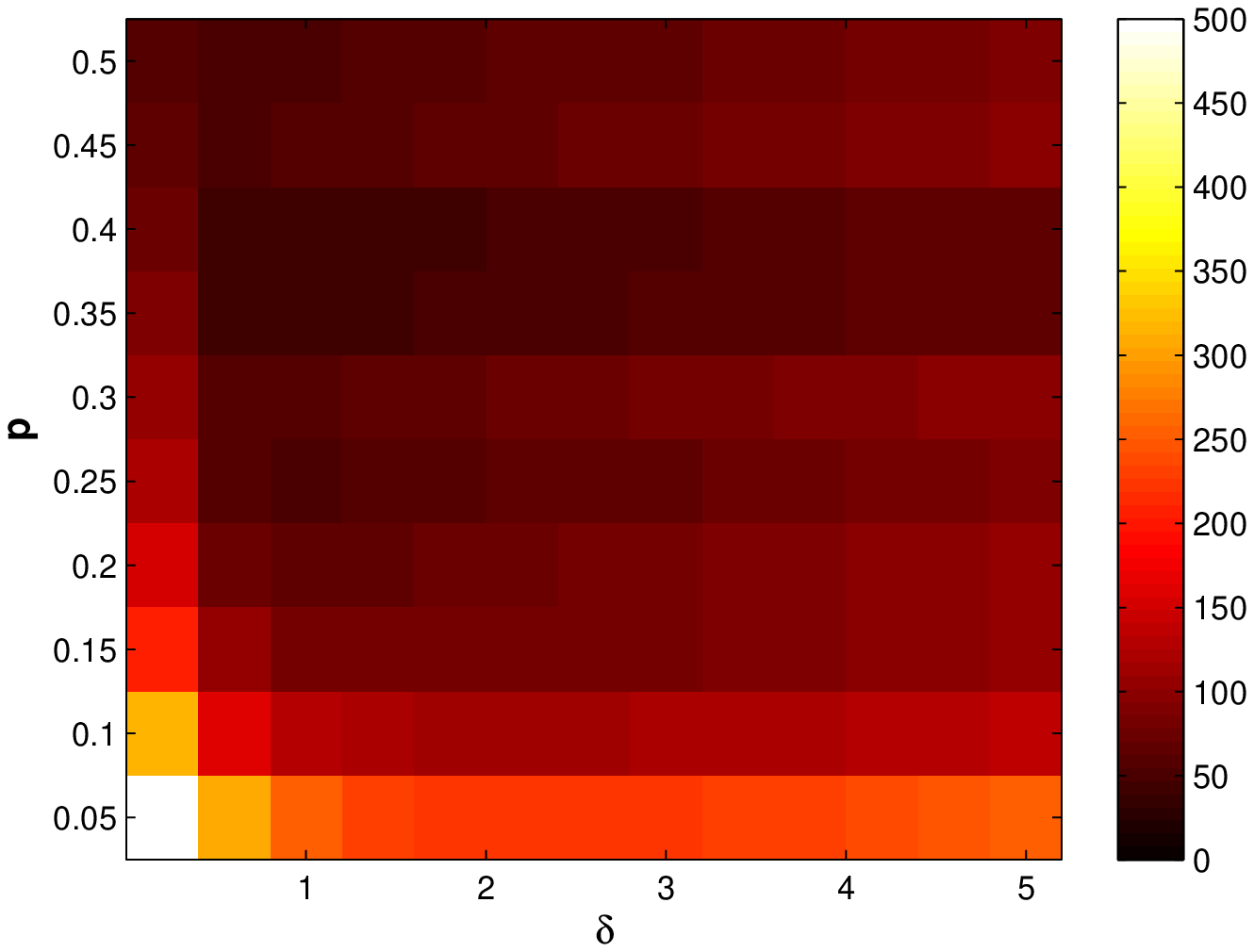,width=9cm}\psfig{figure=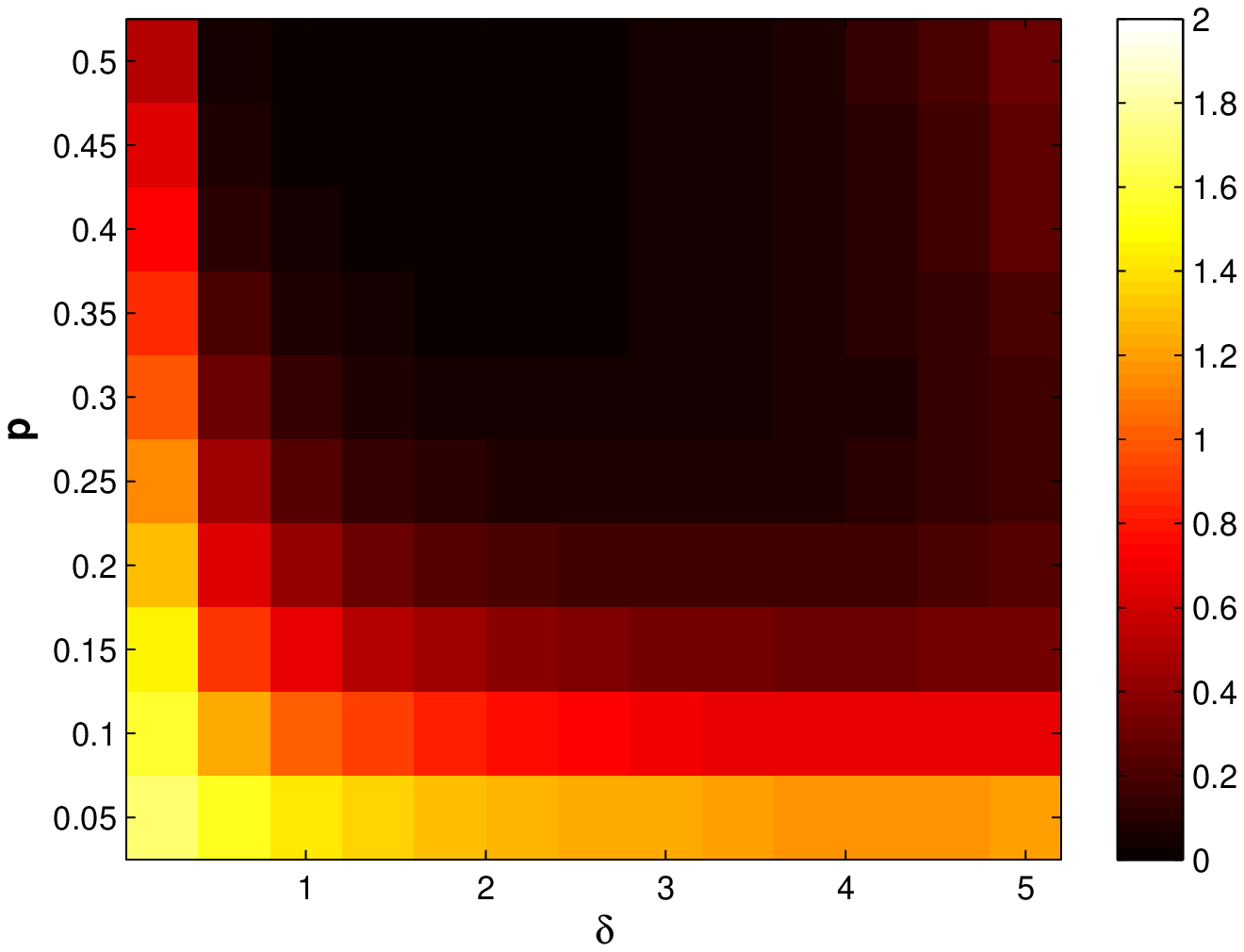,width=9cm}}
\caption{ \small A scale free network of $10^3$ nodes with degree
distribution exponent $\gamma=4$ is considered. The left figure shows
the eigenratio $R$ as varying both the control gain $\delta=\sigma
\kappa$ ($\sigma=0.2$) and the number of controlled nodes (in terms
of the probability $p$).  The right picture shows the control
error at regime $E$ as function of the control gain $\delta= \sigma
\kappa$ and the probability $p$ under the same conditions as in the
left plot. \label{paradoxD34}}
\end{figure}

The main results are shown in Figs.
\ref{paradoxD12},\ref{paradoxD34},
where the theoretical predictions, based on the computation of the
eigenratio $R$, are shown to be pretty well reproduced by the
numerical simulations, involving coupled dynamical systems at the
network vertices.

\begin{figure}[h]
\centerline{
\psfig{figure=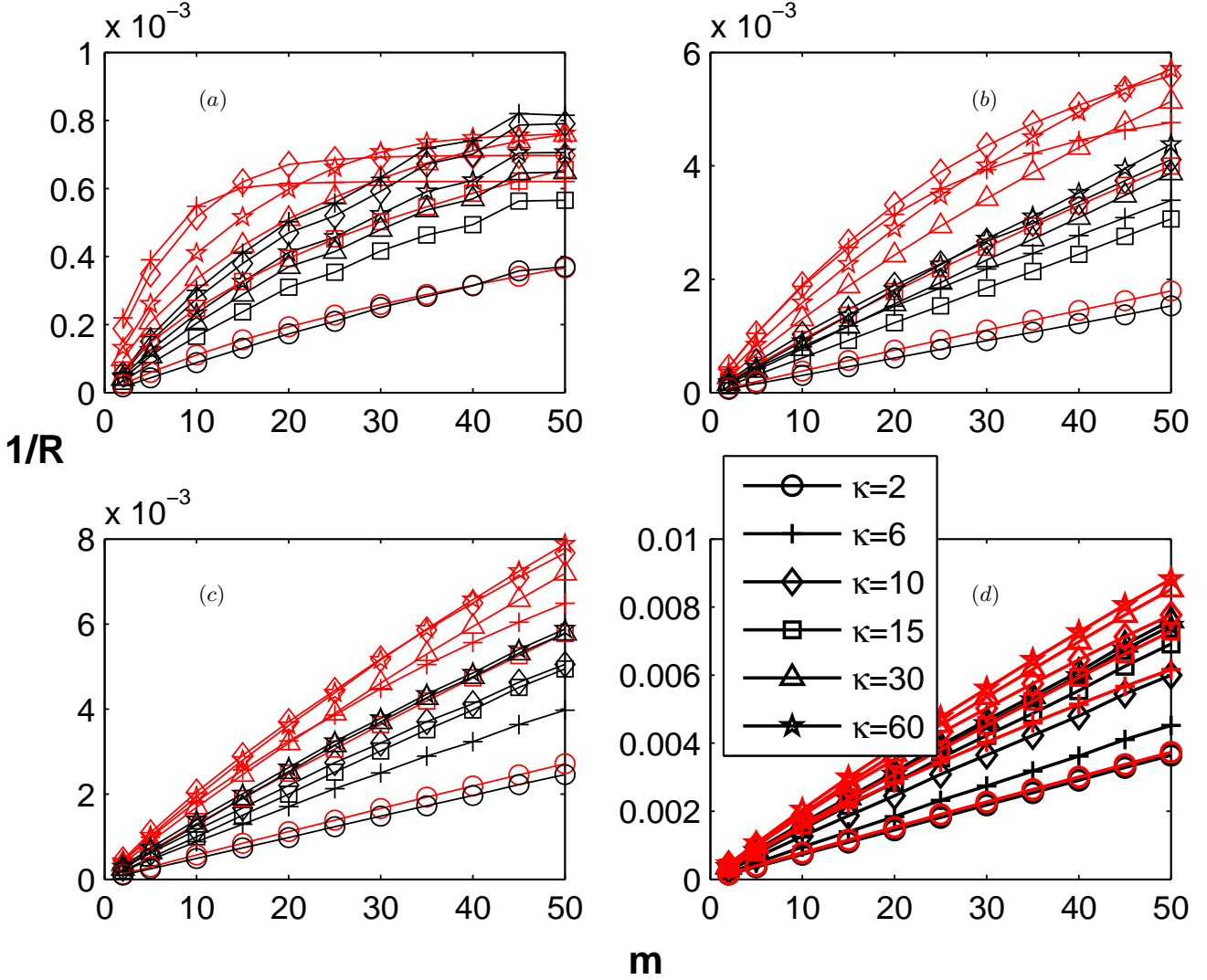,width=20cm}}
\begin{picture}(0,0)(-255,0)
\put(-170,175){\small ${(c)}$} \put(150,175){\small ${(d)}$}
\put(-170,380){\small ${(a)}$} \put(150,380){\small
${(b)}$}\end{picture} \caption{\label{MAIN} \small Networks of
$10^3$ nodes. The controllability $1/R$ is plotted versus the number
of controlled nodes $m$ in networks characterized by different
degree distribution exponents: $\gamma=2.1$ in (a), $\gamma=3$ in
(b), $\gamma=4$ in (c), $\gamma=\infty$ in (d). Black (red) is used
for random (selective) pinning. Different lines represent different
values of the control gain $\kappa$, ranging from $\kappa=2$ to
$\kappa=60$.\label{paradox4}}
\end{figure}

A comparison between random and selective pinning (where the
controlled nodes have been chosen in order of decreasing degree) is
also shown in Fig. \ref{paradox4}. In each subplot
networks characterized by different degree distribution exponents are represented
(in Fig. \ref{paradox4}(d) the network topology is characterized by
an exponential decay of the degree distribution, i.e.
$\gamma=\infty$).

Fig. \ref{paradox4} deserves detailed comments, as listed below:

(i) Notice again that (observe the different scales on the
$y$ axis), networks characterized by higher values of $\gamma$
are more controllable.  The value of the controllability index $1/R$
is particularly low in the case of highly heterogeneous networks
(plot \ref{paradox4}(a)). Specifically, in such a case, its increase
is shown to saturate for high values of $m$, indicating that these
networks are particularly hard to control (i.e., further increase in
the number of controlled nodes does not lead to further improvements
in the network controllability).

(ii) By comparing the results in each subplot, one can observe that
increasing the number of pinned nodes $m$, always results in an
increased controllability, while by varying $\kappa$ a more complex
behavior emerges. Typically, as $\kappa$ grows, first an increase
and then a decrease of $1/R$ is observed, indicating the existence
of optimal ranges of values of $\kappa$ in terms of the network
controllabilty (see also Figs.
\ref{paradox1},\ref{paradoxD12},\ref{paradoxD34}).
These results confirm those previously obtained in \cite{PC} for a
Barabasi-Albert network characterized by degree distribution, $P(k)
\sim k^{-3}$.

(iii) Fig. \ref{paradox4} also shows a comparison between random
strategies (black lines) and selective strategies (red lines) in
choosing the reference sites, where the latter are generally
observed to lead to enhanced network controllability. An exception
is represented by highly heterogeneous networks, in the case where
the number of controlled nodes is sufficiently large. Specifically
in such a case, random strategies can be observed to eventually
outperform selective strategies (see, e.g., the cases of $\kappa=6$
and $\kappa=10$ in Fig.\ref{paradox4}(a)).

\section{Effects of degree correlation}

\begin{figure}[!b]
\begin{center}
\begin{picture}(0,0)(-255,0)
\put(-350,-165){\small ${(d)}$} \put(-350,-40){\small ${(a)}$}
\put(-250,-165){\small ${(e)}$}\put(-150,-165){\small ${(f)}$}
\put(-250,-40){\small ${(b)}$} \put(-150,-40){\small
${(c)}$}\end{picture}
\centerline{\psfig{figure=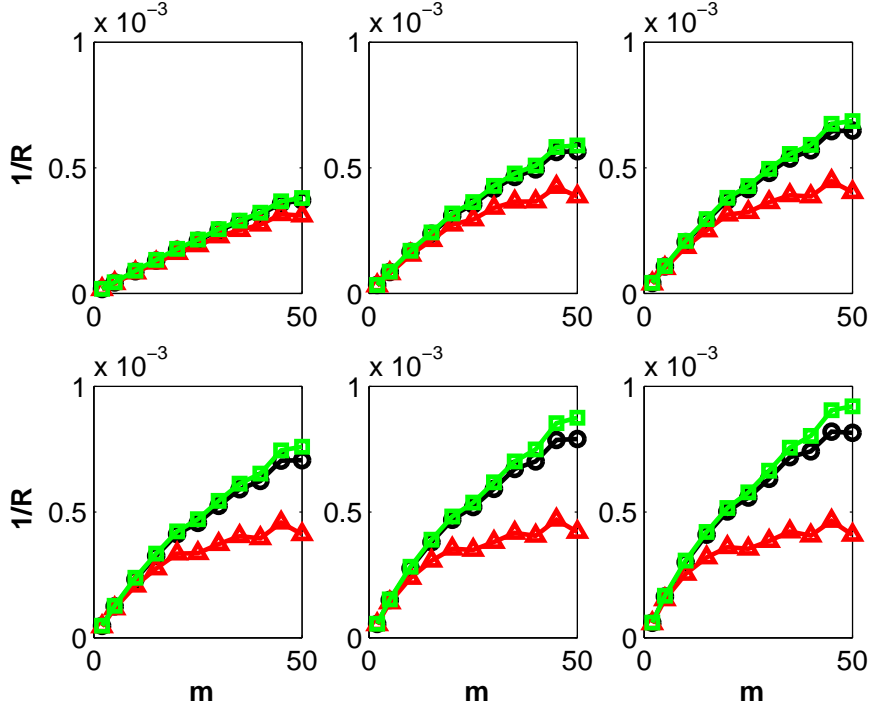,width=13cm}}
\caption{Random pinning of networks characterized by different
degree correlation properties, $N=10^3$, $\mathcal{E}= 4 \cdot
10^3$, $\theta=10$. Here we consider a scale-free heterogeneous
network with degree distribution exponent $\gamma=2.1$. The
controllability is reported versus the number of controlled nodes
$m$, by having in each picture a different value of the control gain
$\kappa$: (a) $\kappa=2$, (b) $\kappa=6$, (c) $\kappa=10$, (d)
$\kappa=15$, (e) $\kappa=30$, (f) $\kappa=50$. Different degree
correlation properties are indicated by different symbols: Triangles
are used in the case of assortative networks ($r=0.3$), squares in
the case of disassortative networks ($r=-0.3$) and circles in the
case of networks that do not display degree correlation ($r=0$).
\label{degreecorr1}}
\end{center}
\end{figure}

\begin{figure}[!t]
\begin{center}
\begin{picture}(0,0)(-255,0)
\put(-350,-165){\small ${(d)}$} \put(-350,-40){\small ${(a)}$}
\put(-250,-165){\small ${(e)}$}\put(-150,-165){\small ${(f)}$}
\put(-250,-40){\small ${(b)}$} \put(-150,-40){\small
${(c)}$}\end{picture}
\centerline{\psfig{figure=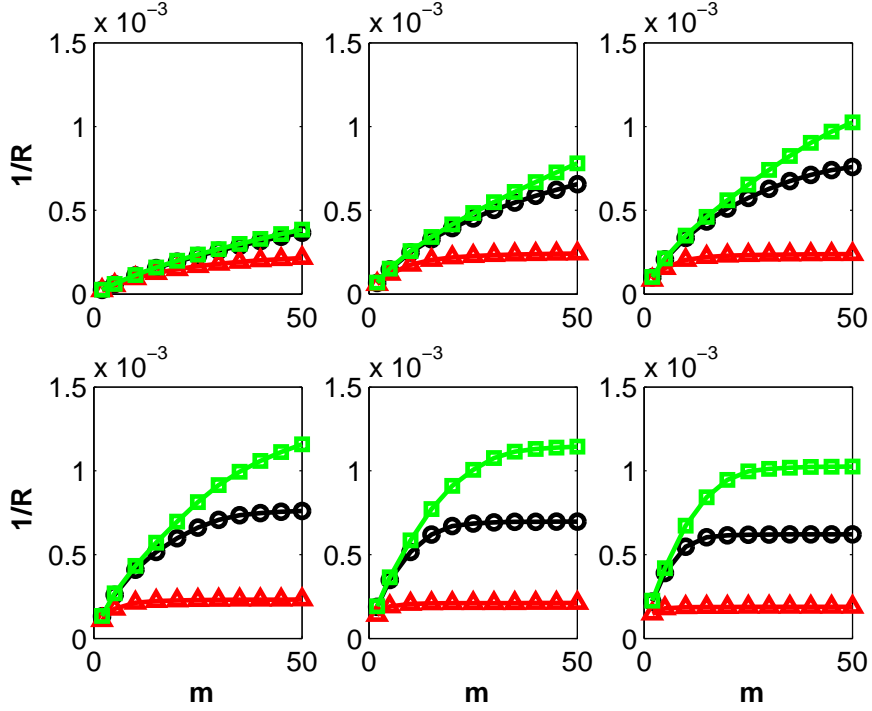,width=13cm}}
\caption{Selective pinning of networks characterized by different
degree correlation properties, $N=10^3$, $\mathcal{E}= 4 \cdot
10^3$, $\theta=10$. Here we consider a scale-free heterogeneous
network with degree distribution exponent $\gamma=2.1$. The
controllability is reported versus the number of controlled nodes
$m$, by having in each picture a different value of the control gain
$\kappa$: (a) $\kappa=2$, (b) $\kappa=6$, (c) $\kappa=10$, (d)
$\kappa=15$, (e) $\kappa=30$, (f) $\kappa=50$. Different degree
correlation properties are indicated by different symbols: Triangles
are used in the case of assortative networks ($r=0.3$), squares in
the case of disassortative networks ($r=-0.3$) and circles in the
case of networks that do not display degree correlation ($r=0$).
\label{degreecorr2}}
\end{center}
\end{figure}

In the preceding section, it was shown that the degree distribution
is indeed an important property in affecting the network
controllability. On the other hand, many other distinctive
properties have been uncovered to characterize in more detail the
structure of real networks, such as, for example, the formation of
communities of strongly interconnected nodes, frequently detected in
many real networks \cite{Ne:Gi02} (for a discussion on the effects
of community structure on the network controllability, see Sec.
VI), or particular forms of correlation or mixing among the network
vertices \cite{New03Mix}.

One form of mixing is the correlation among pairs of linked nodes
according to some properties at the network nodes. A very simple
case is degree correlation \cite{New02Ass}, in which vertices choose
their neighbors according to their respective degrees. Nontrivial
forms of degree correlation have been experimentally detected in
many real-world networks, with social networks being typically
characterized by assortative mixing (which is the case when vertices
are more likely to connect to other vertices with approximately the
same degree) and technological and biological networks, by
disassortative mixing (which takes place when connections are more
frequent between vertices of different degrees). In \cite{New02Ass}
this property has been conveniently measured by means of a single
normalized index, the Pearson statistic $r$ defined as follows:
\begin{equation}
\label{eq:r} r={1
\over{\sigma^2_q}}{\sum_{k,k'}kk'(e_{kk'}-q_k q_{k'})},
\end{equation}
where $q_k$ is the probability that a randomly chosen edge is
connected to a node having degree $k$; $\sigma_q$ is the standard
deviation of the distribution $q_k$ and $e_{kk'}$ represents the
probability that two vertices at the endpoints of a generic edge
have degrees $k$ and $k'$, respectively. Positive values of $r$
indicate assortative mixing, while negative values characterize
disassortative networks.

The effects of degree correlation on the network synchronizability
have been studied in \cite{So:di:Bo}. Specifically in
\cite{So:di:Bo}, the network synchronizability  has been shown to be
enhanced as the network becomes more disassortative (i.e. $r$
decreases) for both systems in class I and II (namely, what is
observed is that the second smallest eigenvalue varies sensibly with
$r$, while the largest one is weakly influenced by variations of the
network degree correlation). In what follows, we attempt to
characterize the effects of degree correlation on the network
controllability.

Specifically, by following a strategy similar to the one presented
in \cite{New02Ass,reshuffling}, which allows to vary the degree
correlation (in terms of variable values of $r$) while keeping
the degree distribution fixed, we show that disassortative mixing,
i.e. the tendency of high-degree nodes to establish connections with
low-degree ones (and viceversa), is indeed a desirable network
property in terms of its controllability.

The main results are shown in Figs. \ref{degreecorr1} and
\ref{degreecorr2}. Note that in all the cases considered (many
simulations were carried out involving different values of the
control gain $\kappa$, different numbers of controlled nodes $m$,
chosen according to different strategies, and different degree
distribution exponents $\gamma$), negative degree correlation has
always been found to enhance the network controllabilty.

Moreover the effects of degree correlation seem to be strongly
emphasized when large gains $\kappa$ are used to control the network
and selective pinning strategies are considered. For example, in the
cases represented in Fig. \ref{degreecorr2}(d-f), where selective
pinning is used in combination with large control gains at the
pinned nodes, the variations in the controllability for networks
characterized by different degree-degree mixing, are impressive.

Another interesting phenomenon is observed when different pinning
strategies are used to control assortatively mixed networks. In
fact, as can be observed by comparing Figs.\ref{degreecorr2}(d-f)
and \ref{degreecorr1}(d-f), under certain conditions, random pinning
results more effective than selective pinning in controlling such
networks.

As we will show in the following, 
this surprising phenomenon can be explained in terms of the
distribution of the controllers over the whole network.
Specifically, when the controllers are located at the high degree
nodes (selective pinning) and these are all linked together
(assortative mixing), this causes a loss of their capability to
control the rest of the network. Therefore in such a case,
random pinning (i.e. more uniform distribution of the controllers)
turns out to be more effective than selective pinning. On the other hand,
when the controllers are located at the high degree nodes and these
can maximize their influence being connected to many low degree ones
(disassortative mixing), we observe that the network controllability
is strongly enhanced.

%
%

\section{Linear and square lattices}

\begin{figure}[h]
\begin{center}
\psfig{figure=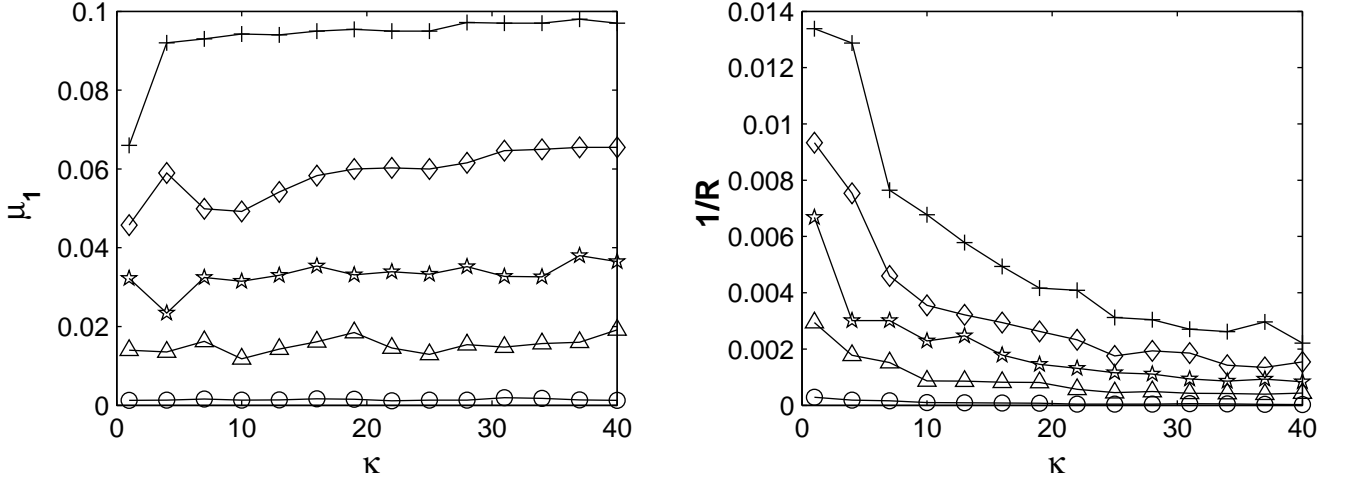}
\end{center}
\caption{A monodimensional lattice of $10^3$ nodes is considered.
The controllability indices $\mu_1$ and $1/R$ are plotted as
functions of the control gain $\kappa$.  The legend is as follows:
$p=0.05$ (circles), $p=0.20$ (triangles), $p=0.30$ (stars), $p=0.40$
(diamonds), $p=0.50$ (plus).\label{mono}}
\end{figure}

As an example of a very simple network, we consider here a linear
(monodimensional) lattice of $N$ nodes.  Each node $i=2,...,N-1$ is
connected to nodes $i-1$ and $i+1$; we assume periodic boundary
conditions, i.e. node $1$ is connected to nodes $N$ and $2$, and node
$N$ to nodes $N-1$ and $1$. The controllability of such a network is
shown in Fig. \ref{mono}, as varying both the number of controlled
nodes $m$ and the control gain $\kappa$. As Fig.
\ref{mono}(left plot) shows, using a larger number of
controllers is effective in enhancing the network controllability $\mu_1$, while
only a little improvement is experimented by increasing the control
gain $\kappa$. On the other hand, increasing the control gain
$\kappa$, leads to a sharp decrease of the controllability $1/R$ for
the dynamical systems in class I (as shown in the right-hand side
plot).

\begin{figure}[h]
\centerline{ \psfig{figure=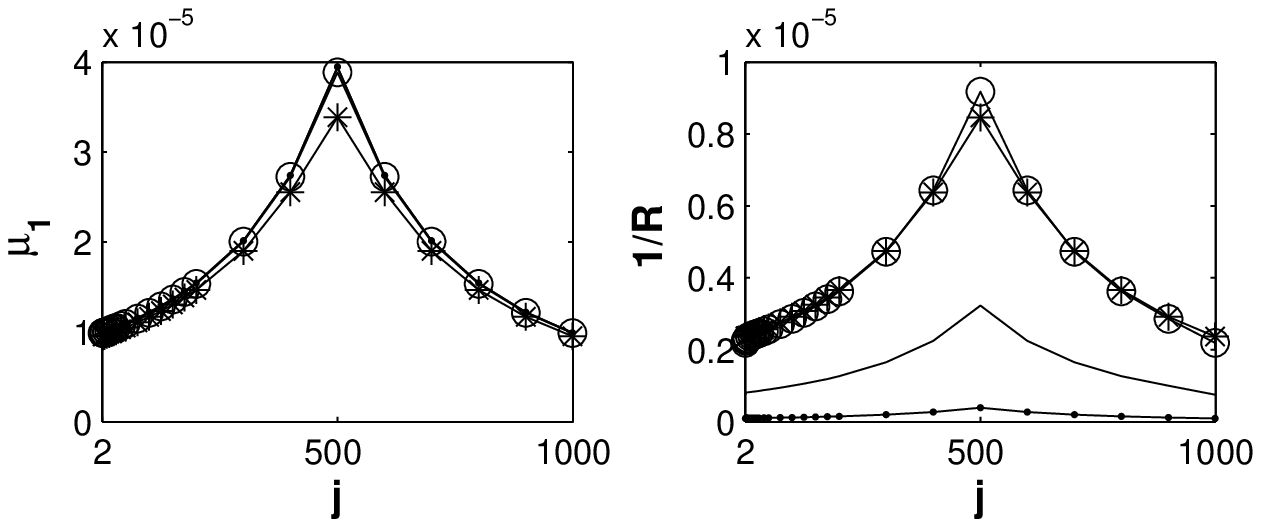,width=18cm}}
\caption{A
monodimensional lattice of $10^3$ nodes is considered. Assuming that
the controlled nodes are node $1$ and node $j$, the controllability
indices $\mu_1$ and $1/R$ are plotted versus $j$. Different lines
indicate different control gains $\kappa$: $\kappa=10^{-1}$
(asterisks), $\kappa=10^0$ (circles), $\kappa=10^1$ (no marker),
$\kappa=10^2$ (dots). \label{monod}}
\end{figure}

In what follows, we address the issue of which is the best
combination of nodes to control this simple model of
network,
in order to increase its controllability. Imagine one has already
selected one node $i$ at random, say $i=1$ (without any loss of
generality). We now wonder which node $j$ should be selected
then, to maximize the network controllability. In Fig. \ref{monod},
it is shown how the network controllability is affected by the
choice of the second node $j=2,...,N$. Interestingly, the best
choice is performed when the furthest node from $i$ is selected,
i.e. $j=501$. Note that  this result is not influenced by the
particular value of the control gain $\kappa$ (as shown in Fig.
\ref{monod}).

Furthermore, we have considered more complex situations in which three
or more nodes were to be selected. Interestingly, the combination
that optimizes the network controllability is always the one that
maximizes the average distance among the selected nodes. Hence, we conjecture 
that the distance among the selected nodes to be controlled in
a network is an important aspect one should consider, when designing
pinning control schemes.

As a further example, we consider the case of a square lattice with
periodic boundary conditions. Lattices are structured regular
networks in which all the nodes have the same degree (also termed as
the coordination number $z$) and are widely used in physics to
describe phenomena that take place in ordered extended systems.
For example, pinning control of coupled map lattices has been
studied in \cite{Pinn1,Pinn2}.

In what follows, we assume the network takes the form of a square
lattice, consisting of $N=l \times l$ nodes, where $l$ is the side
of the lattice.  Each node $i$ can be mapped 
into a point of integer
coordinates ($x^i,y^i$), with $x^i=1,...,l$ and $y^i=1,...,l$, every site being 
linked to its $z=4$ nearest neighbors. We assume periodic boundary
conditions and hence the network can be seen as having a toroidal
topology in which nodes on the edge of the lattice are connected to
those on the opposite edge. This is also known as a
\textit{Manhattan} lattice.

\begin{figure}[h]
\begin{center}
\psfig{figure=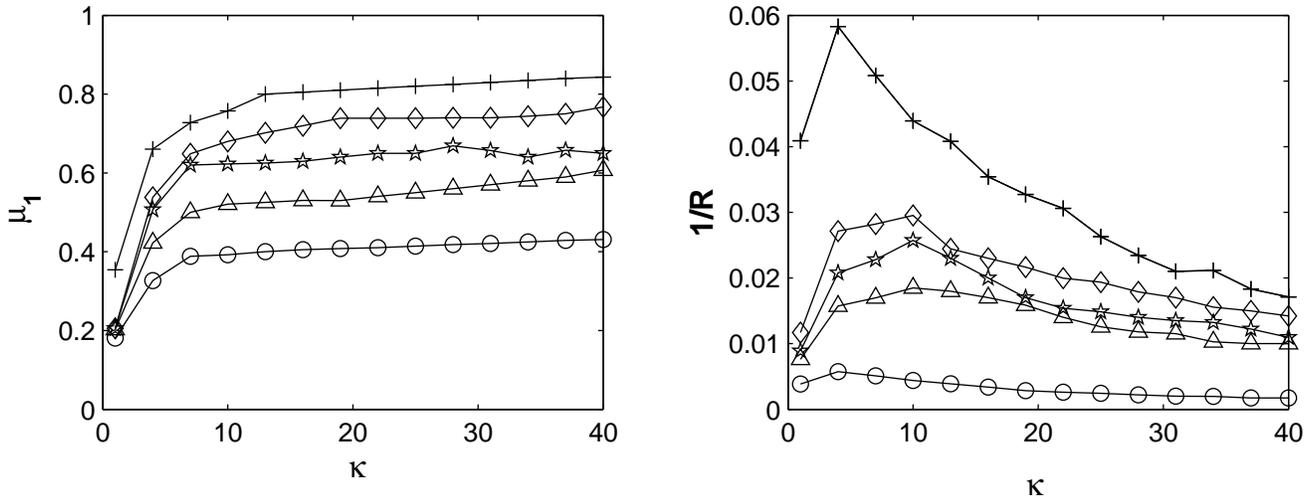}
\end{center}
\caption{A periodic square lattice of $32 \times 32$ nodes ($z=4$)
is considered. The controllability indices $\mu_1$ and $1/R$ are
plotted as functions of the control gain $\kappa$.  The legend is as
follows: $p=0.05$ (circles), $p=0.20$ (triangles), $p=0.30$ (stars),
$p=0.40$ (diamonds), $p=0.50$ (plus). \label{square}}
\end{figure}

The controllability for such a network is shown in Fig.
\ref{square}, where both $\mu_1$ and $1/R$ are plotted versus the
control gain $\kappa$. Note that there are intermediate values of
the control gain $\kappa$ which maximize $1/R$ (as shown in the left
panel of Fig. \ref{square}). Thus, either too large or too small
values of $\kappa$ can reduce the controllability of square
lattices.

Again, as in the preceding case, we are interested in the best strategy to control
the lattice, once a given number of controlled nodes (e.g., two) is
given. To this aim, a node is selected at random from the network,
say $i$ (whose choice will not affect our results, due to the
isotropy property of the lattice). Then the other nodes in the network are grouped according to their distance from $i$,
which ranges between $1$ and $l$. Note that this time, the choice of
which node is selected among all those at a certain distance $d$
from $i$ is not ineffective in terms of the network
controllability. However, we have checked the variations (in terms of
controllability) among all the nodes at a certain distance from $i$
to be negligible when compared to those among nodes at different
distances.

\begin{figure}[h]
\centerline{ \psfig{figure=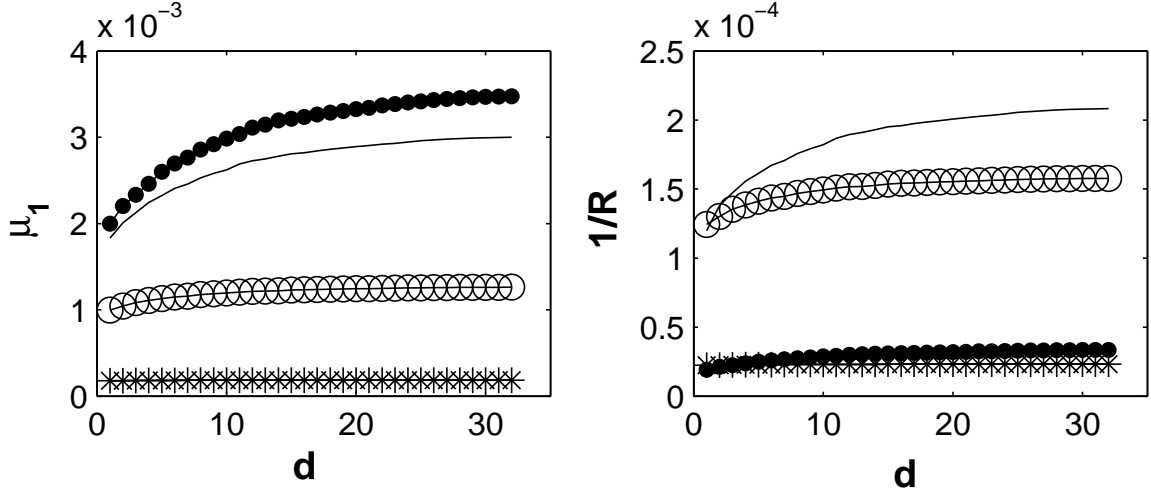,width=18cm}}
\caption{A square
lattice of $32 \times 32$ nodes is considered. Assuming that two of
its nodes have been chosen to be controlled, the controllability
indices $\mu_1$ and $1/R$ are plotted versus their distance $d$.
Different lines indicate different control gains $\kappa$:
$\kappa=10^{-1}$ (asterisks), $\kappa=10^0$ (circles), $\kappa=10^1$
(no marker), $\kappa=10^2$ (dots). \label{squad}}
\end{figure}

The results are shown in Fig. \ref{squad}, where both $\mu_1$ and
$1/R$ are reported, 
as varying the distance $d$ between the controlled nodes $i$ and $j$
(in the figure, each point has been averaged over different choices of the node $j$
at a given distance $d$ from $i$). Again the best case in terms of
controllability is obtained when $d$ is maximized.

This confirms our previous conjecture, that the distribution of the
controlled nodes over the network is indeed an important property
one should consider when designing pinning control schemes. In other
words, the more uniformly the nodes are distributed over the
network, the higher is their ability to control it.

Previous studies \cite{PinnA,PC} have already considered different
strategies (i.e. random vs. selective strategies) in selecting the
best suited reference sites over a given complex network. Therein,
it has been proposed that the degree of the selected vertices is
indeed an important factor. Here we propose that the uniformity of
the distribution of the reference sites over the network (in terms
e.g., of the total distance among them) is an important aspect as
well. In the next section, we will seek to provide further evidence
of this simple general principle, by considering the controllability
of complex networks characterized by community structure.

\section{Community Structure}

Synchronizability and synchronization dynamics of networks
characterized by community structure have been previously studied in
\cite{Oh:Rho,Ar:Di:PV,Pa:Lai}. In \cite{Oh:Rho}, the interplay
between modular synchronization and global synchronization has been
discussed for networks affected by community structure. In
\cite{Ar:Di:PV}, the dynamic time scales of hierarchical
synchronization among network communities have been studied, and a
connection between the spectral information of the whole Laplacian
spectrum and the dynamical process of modular synchronization has
been reported.

Here we shall seek to analyze the controllability of networks
affected by community structure \cite{Ne:Gi02}.  Community structure
arises when the number of edges between vertices belonging to
different communities is considerably smaller than the number of
edges falling inside of the communities  \footnote{Actually it would
be more correct to affirm that community structure is detected when
the number of connections falling between different communities is
considerably lower than one would expect by considering another
network characterized by the same degree distribution but completely
random with respect to any other aspect \cite{Ne:Gi04}.}.

Next we propose a dynamical process, that starting from a generic
network configuration and an arbitrary subdivision of its
nodes in communities, allows to strengthen the network community 
structure. Specifically, this is achieved by having the
inter-community connections dynamically rewired inside of the
single communities.

Let us start with a given general complex network and 
assume a random subdivision of the networks nodes in two
communities, say $C_1$ and $C_2$. Say $N_1$ the number of nodes in
$C_1$ and $N_2=N-N_1$ the number of nodes in $C_2$. Consider the set
$S$ of all the connections falling between $C_1$ and $C_2$ and say
$s$ its cardinality. At the beginning the number of links in $S$ is
$s_0$. At every single step of the algorithm, the following
operations are performed:

1) A pair of links is selected from $S$. Let us denote by $v_1$ and
$w_1$ the end points of the two selected edges in $C_1$, and $v_2$
and $w_2$ their end points in $C_2$.

2) In the case in which $v_1$ and $w_1$ (and $v_2$ and $w_2$) are not
already  connected, the selected links are eliminated from the
network and two new connections, namely ($v_1$,$w_1$) and
($v_2$,$w_2$) are created.

3) In the case in which the switch in 2) is successfully performed, $s$
is decreased by 2.

Note that the algorithm described above is ergodic over the degree
distribution \cite{New03Mix}.

Hereafter, for the sake of simplicity, we will consider a network
generated by the algorithm described in Sec. III, characterized by
exponential decay of the degree distribution (i.e., $\gamma=\infty$).
One-fourth of the nodes are then randomly selected to form the first
community $C_1$. The remaining ones are those in $C_2$. We set
the pinning probability to $p=0.1$ and randomly choose the reference
sites among all the networks nodes.

\begin{figure}[h]
\centerline{ \psfig{figure=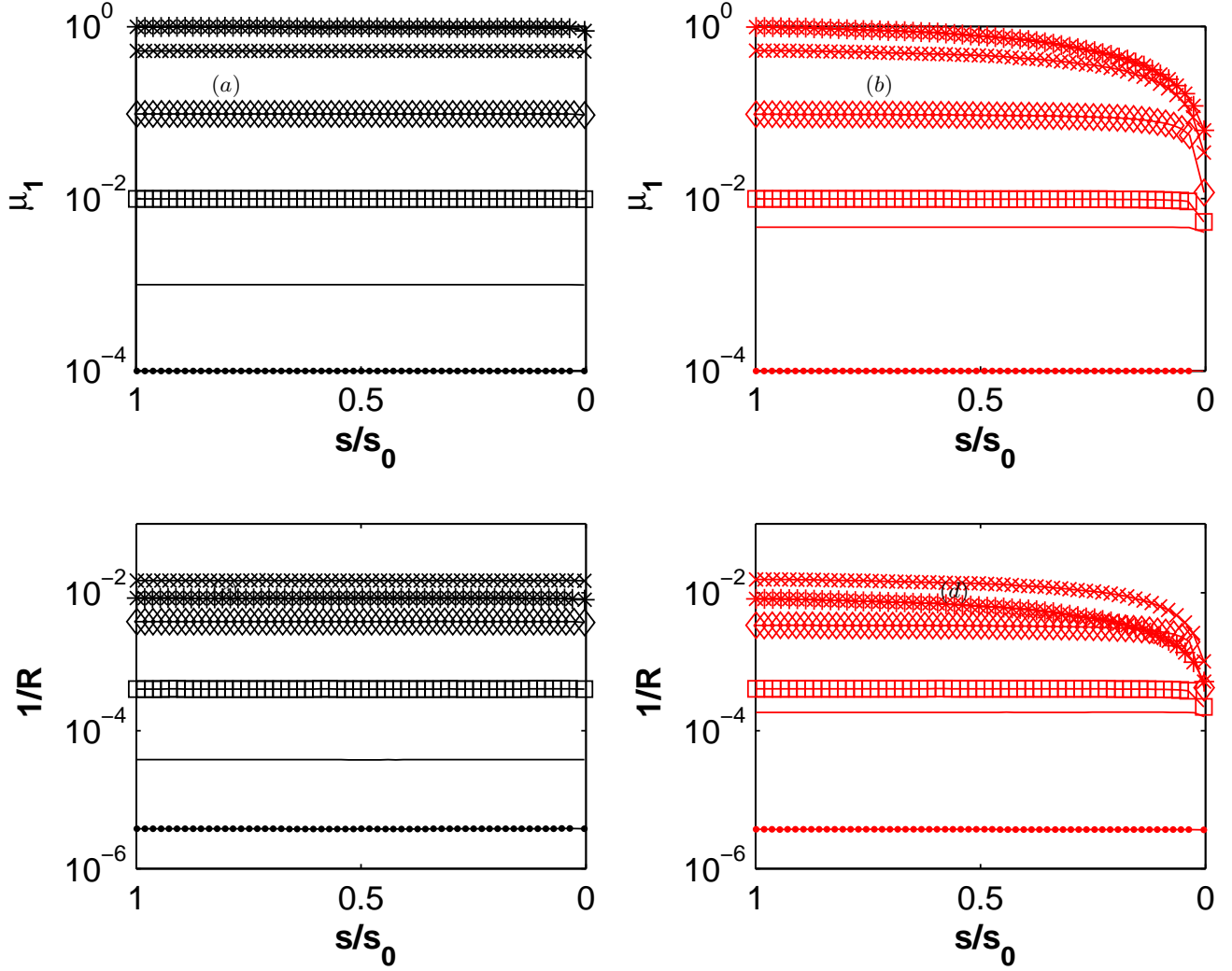,width=20cm}}
\begin{picture}(0,0)(-255,0)
\put(-170,175){\small ${(c)}$} \put(125,175){\small ${(d)}$}
\put(-170,380){\small ${(a)}$} \put(95,380){\small
${(b)}$}\end{picture}\caption{A random network of $5 \cdot 10^2$
nodes composed of two communities is considered. The first one
$C_1$, includes $125$ nodes and the second one $C_2$, includes the
remaining ($375$) ones. The network controllability $\mu_1$ ($1/R$)
is plotted versus $\frac{s}{s_0}$ in (a) and (b) ((c) and (d)) as
$s$ is made dynamically vary between $1$ and $0$, according to the
technique described in the text. The pinning probability is assumed
to be $p=0.1$. Different lines indicate several values of the
control gain $\kappa$: $\kappa=10^{-3}$ (points), $\kappa=10^{-2}$
(no marker), $\kappa=10^{-1}$ (squares), $\kappa=10^{0}$ (diamonds),
$\kappa=10^{1}$ (times), $\kappa=10^{2}$ (asterisks).  The left
plots (a) and (c), show the cases where the controllers are evenly
distributed among the two communities. The right plots (b) and (d)
show a situation where the $95$ percent of the controllers belong to
$C_1$ and the remaining to $C_2$. Observe the logarithmic scale on
the $y$ axis. \label{CSI}}
\end{figure}

In Figs. \ref{CSI}(a) and \ref{CSI}(c), the behavior of the network
controllability is reported as $s/s_0$ is varied
according to the iterative process described above (note that in so doing, only the connections among the network nodes are rewired, without
modifying the choice of the reference sites).

The figures show high sensitivity of the controllability with
respect to the choice of the control gain $\kappa$, ranging between
$10^{-3}$ and $10^2$. However, neither $\mu_1$ nor $1/R$ is
influenced by the dynamical links redirection process proposed in
this section. Even in the limit case in which $s=0$, which corresponds
to the network being disconnected in two separate clusters, its
controllability is not observed to go to $0$ (however this is true
as long as at least one controller is present in each of the
isolated clusters).
It is worth noting that this represents another main difference
between control and synchronization of complex networks.
Specifically, in order to control a network, it is not necessary to
have a unique globally connected cluster. Namely, this condition can
be replaced by the one, that states that in each isolated cluster, at least one controller is present.

So far, we have assumed the reference sites to be uniformly randomly
selected from the set of all the network nodes (i.e. both $C_1$ and
$C_2$). A different situation is instead shown in the right panels
of Figs. \ref{CSI}, where we have assumed that about $95$ per cent
of the reference sites belong to $C_1$ and only the remaining $5$
per cent to $C_2$. Similarly to Sec. V, we do this with the aim of unveiling how the
distribution of the reference sites
among the network nodes 
affects its controllability (but here, different from Sec. V, we consider complex random networks). In so doing, we compare the controllability of modular
networks according to the distribution of the controllers among their
different communities.
Furthermore, this is motivated by real networks observation. For
instance, the control of the synchronous beat of the heart cells is exerted by the pacemaker
cells, which are grouped together at the sinoatrial node.

The simulations in the right panels of Fig. \ref{CSI} clearly show
that in this case, as $s$ is dynamically reduced from $1$ to $0$,
both $\mu_1$ and $1/R$ are influenced by
the emergence of the network community structure. In particular, 
the non-uniform distribution of the reference sites over the
network, reduces the network controllability.

 Loosely speaking,
this is because the number of reference sites in one of the
communities (i.e. $S_2$) is too low, so that this part of
the network is \emph{out of control}. 
In other words, as it is clearly shown in Figs. \ref{CSI} (a-d), the
emergence of clusters of reference sites is not a desirable property
in terms of the network controllability.

In the case of the heartbeat control, we conjecture that the cluster
organization of the pacemakers, might be explained in terms of the
self-sustained oscillations generated by the interactions among the
pacemakers themselves. Thus in such a case, one should consider (i)
the influence of the reference sites on the other nodes, (ii) the
dynamics among the non-controlled nodes, and also (iii) the dynamics
occurring among the pacemakers, as well. 
Note that in the case of the heartbeat, different types of dynamics characterize
the reference and the other sites, as witnessed by the difference in
the phase and the amplitudes of their pulsations. This represents an
interesting subject for the future research activity.

%

\section{Conclusions}

Motivated by the results reported in a recent paper, where a technique
was introduced to
describing the controllability of networks under pinning control
schemes,
in this paper we have compared several networks topologies in terms of their controllability.
Specifically, we have considered networks in which two different layers of
dynamical nodes coexist: the uncontrolled sites and the reference
(controlled) ones, where the latter play the role of
leading the whole network to evolve toward a desired reference
evolution. 

In \cite{PC}, a surprising property was pointed out (for the dynamical
systems in class I), i.e., the network controllability is
reduced when the average control gain $\kappa$ is increased to beyond a
certain value. Interestingly, this property has been confirmed here to be
reproduced over a wide variety of different network topologies and even
lattices.

This finding is likely to offer an explanation to the daily life
experience of everybody that in order to deal with 
complex systems/situations, one should adopt 
the right dose of strength in the control action:
either a too weak or too strong action could lead to a
loss of control over the whole system/situation.

In this paper, we have considered the effects of heterogeneity in
the degree distribution, degree correlation, as well as community
structure on the network controllability. High heterogeneity in the
degree distribution has been found to reduce the network
controllability, while disassortative mixing enhances it. Moreover,
motivated by the observations on simple networks, such as linear and
square lattices, we have conjectured that a uniform distribution of
the reference sites over a network is indeed important regarding
its controllability. This finding could be useful for future control theory of complex networks.

This has been confirmed also when more complex topologies have been
considered. In particular, in the case of assortatively mixed
networks (characterized by positive degree correlation), we have
found that controlling the group of the high degree nodes (which is
commonly believed to enhance the control performance) is indeed an
ineffective strategy. At the same time, in the case of networks
characterized by strong community structure, it has been shown that
placing most of the controllers within the same community can lead to
a loss of control over the whole network.

%
%

Acknowledgements: The author acknowledges help in the writing of the paper from Guanrong Chen. He thanks Franco Garofalo, Mario di Bernardo, Guanrong Chen, Zhengping Fan, Mark Avrum Gubrud and Zeynep Tufekci for useful comments and discussions.


\begin{thebibliography}{32}
\expandafter\ifx\csname
natexlab\endcsname\relax\def\natexlab#1{#1}\fi
\expandafter\ifx\csname bibnamefont\endcsname\relax
  \def\bibnamefont#1{#1}\fi
\expandafter\ifx\csname bibfnamefont\endcsname\relax
  \def\bibfnamefont#1{#1}\fi
\expandafter\ifx\csname citenamefont\endcsname\relax
  \def\citenamefont#1{#1}\fi
\expandafter\ifx\csname url\endcsname\relax
  \def\url#1{\texttt{#1}}\fi
\expandafter\ifx\csname urlprefix\endcsname\relax\def\urlprefix{URL
}\fi \providecommand{\bibinfo}[2]{#2}
\providecommand{\eprint}[2][]{\url{#2}}

\bibitem[{\citenamefont{Sorrentino et~al.}(2007)\citenamefont{Sorrentino,
  di~Bernardo, Garofalo, and Chen}}]{PC}
\bibinfo{author}{\bibfnamefont{F.}~\bibnamefont{Sorrentino}},
  \bibinfo{author}{\bibfnamefont{M.}~\bibnamefont{di~Bernardo}},
  \bibinfo{author}{\bibfnamefont{F.}~\bibnamefont{Garofalo}}, \bibnamefont{and}
  \bibinfo{author}{\bibfnamefont{G.}~\bibnamefont{Chen}}, \bibinfo{journal}{Phys. Rev. E}
  \textbf{\bibinfo{volume}{75}}, \bibinfo{pages}{046103}
  (\bibinfo{year}{2007}).

\bibitem[{\citenamefont{Peskin}(1977)}]{PesBOOK}
\bibinfo{author}{\bibfnamefont{C.}~\bibnamefont{Peskin}},
  \emph{\bibinfo{title}{Mathematical aspects of heart physiology}}
  (\bibinfo{publisher}{Courant Inst. Math Sci., New York},
  \bibinfo{year}{1977}).

\bibitem[{\citenamefont{Yamaguchi et~al.}(2003)\citenamefont{Yamaguchi,
  Isejima, Matsuo, Okura, Yagita, Kobayashi, and Okamura}}]{supra}
\bibinfo{author}{\bibfnamefont{S.}~\bibnamefont{Yamaguchi}},
  \bibinfo{author}{\bibfnamefont{H.}~\bibnamefont{Isejima}},
  \bibinfo{author}{\bibfnamefont{T.}~\bibnamefont{Matsuo}},
  \bibinfo{author}{\bibfnamefont{R.}~\bibnamefont{Okura}},
  \bibinfo{author}{\bibfnamefont{K.}~\bibnamefont{Yagita}},
  \bibinfo{author}{\bibfnamefont{M.}~\bibnamefont{Kobayashi}},
  \bibnamefont{and} \bibinfo{author}{\bibfnamefont{H.}~\bibnamefont{Okamura}},
  \bibinfo{journal}{Science} \textbf{\bibinfo{volume}{302}},
  \bibinfo{pages}{1408} (\bibinfo{year}{2003}).

\bibitem[{\citenamefont{Pecora and Carroll}(1998)}]{Pe:Ca}
\bibinfo{author}{\bibfnamefont{L.}~\bibnamefont{Pecora}} \bibnamefont{and}
  \bibinfo{author}{\bibfnamefont{T.}~\bibnamefont{Carroll}},
  \bibinfo{journal}{Phys. Rev. Lett.} \textbf{\bibinfo{volume}{80}},
  \bibinfo{pages}{2109} (\bibinfo{year}{1998}).

\bibitem[{\citenamefont{Nishikawa et~al.}(2003)\citenamefont{Nishikawa, Motter,
  Lai, and Hoppensteadt}}]{Ni:Mo}
\bibinfo{author}{\bibfnamefont{T.}~\bibnamefont{Nishikawa}},
  \bibinfo{author}{\bibfnamefont{A.}~\bibnamefont{Motter}},
  \bibinfo{author}{\bibfnamefont{Y.}~\bibnamefont{Lai}}, \bibnamefont{and}
  \bibinfo{author}{\bibfnamefont{F.}~\bibnamefont{Hoppensteadt}},
  \bibinfo{journal}{Phys. Rev. Lett.} \textbf{\bibinfo{volume}{91}},
  \bibinfo{pages}{014101} (\bibinfo{year}{2003}).

\bibitem[{\citenamefont{Hwang et~al.}(2005)\citenamefont{Hwang, Chavez, Amann,
  and Boccaletti}}]{Bocc2}
\bibinfo{author}{\bibfnamefont{D.}~\bibnamefont{Hwang}},
  \bibinfo{author}{\bibfnamefont{M.}~\bibnamefont{Chavez}},
  \bibinfo{author}{\bibfnamefont{A.}~\bibnamefont{Amann}}, \bibnamefont{and}
  \bibinfo{author}{\bibfnamefont{S.}~\bibnamefont{Boccaletti}},
  \bibinfo{journal}{Phys. Rev. Lett.} \textbf{\bibinfo{volume}{94}},
  \bibinfo{pages}{138701} (\bibinfo{year}{2005}).

\bibitem[{\citenamefont{Boccaletti et~al.}(2006)\citenamefont{Boccaletti,
  Latora, Moreno, Chavez, and Hwang}}]{report}
\bibinfo{author}{\bibfnamefont{S.}~\bibnamefont{Boccaletti}},
  \bibinfo{author}{\bibfnamefont{V.}~\bibnamefont{Latora}},
  \bibinfo{author}{\bibfnamefont{Y.}~\bibnamefont{Moreno}},
  \bibinfo{author}{\bibfnamefont{M.}~\bibnamefont{Chavez}}, \bibnamefont{and}
  \bibinfo{author}{\bibfnamefont{D.}~\bibnamefont{Hwang}},
  \bibinfo{journal}{Physics Reports} \textbf{\bibinfo{volume}{424}},
  \bibinfo{pages}{175} (\bibinfo{year}{2006}).

\bibitem[{\citenamefont{Strogatz}(2003)}]{SYNCBOOK}
\bibinfo{author}{\bibfnamefont{S.}~\bibnamefont{Strogatz}},
  \emph{\bibinfo{title}{Sync: The Emerging Science of Spontaneous Order}}
  (\bibinfo{publisher}{Hyperion. New York}, \bibinfo{year}{2003}).

\bibitem[{\citenamefont{Czeisler et~al.}(1986)\citenamefont{Czeisler, Allan,
  Strogatz, Ronda, Sanchez, Rios, Freitag, Richardson, and Kronauer}}]{bright}
\bibinfo{author}{\bibfnamefont{C.~A.} \bibnamefont{Czeisler}},
  \bibinfo{author}{\bibfnamefont{J.~S.} \bibnamefont{Allan}},
  \bibinfo{author}{\bibfnamefont{S.~H.} \bibnamefont{Strogatz}},
  \bibinfo{author}{\bibfnamefont{J.~M.} \bibnamefont{Ronda}},
  \bibinfo{author}{\bibfnamefont{R.}~\bibnamefont{Sanchez}},
  \bibinfo{author}{\bibfnamefont{C.~D.} \bibnamefont{Rios}},
  \bibinfo{author}{\bibfnamefont{W.~O.} \bibnamefont{Freitag}},
  \bibinfo{author}{\bibfnamefont{G.~C.} \bibnamefont{Richardson}},
  \bibnamefont{and} \bibinfo{author}{\bibfnamefont{R.~E.}
  \bibnamefont{Kronauer}}, \bibinfo{journal}{Science}
  \textbf{\bibinfo{volume}{233}}, \bibinfo{pages}{667} (\bibinfo{year}{1986}).

\bibitem[{\citenamefont{McClintock}(1971)}]{McC71}
\bibinfo{author}{\bibfnamefont{M.}~\bibnamefont{McClintock}},
  \bibinfo{journal}{Nature} \textbf{\bibinfo{volume}{229}},
  \bibinfo{pages}{244} (\bibinfo{year}{1971}).

\bibitem[{\citenamefont{Russell and abd K.~Thomson}(1980)}]{Rus80}
\bibinfo{author}{\bibfnamefont{M.}~\bibnamefont{Russell}} \bibnamefont{and}
  \bibinfo{author}{\bibfnamefont{G.~S.} \bibnamefont{abd K.~Thomson}},
  \bibinfo{journal}{Pharmacol. Biochem. Behav.} \textbf{\bibinfo{volume}{13}},
  \bibinfo{pages}{737} (\bibinfo{year}{1980}).

\bibitem[{\citenamefont{Kori and Mikhailov}(2006)}]{Mikh1}
\bibinfo{author}{\bibfnamefont{H.}~\bibnamefont{Kori}} \bibnamefont{and}
  \bibinfo{author}{\bibfnamefont{A.~S.} \bibnamefont{Mikhailov}},
  \bibinfo{journal}{Phys. Rev. E} \textbf{\bibinfo{volume}{74}},
  \bibinfo{pages}{066115} (\bibinfo{year}{2006}).

\bibitem[{\citenamefont{Kori and Mikhailov}(2004)}]{Mikh2}
\bibinfo{author}{\bibfnamefont{H.}~\bibnamefont{Kori}} \bibnamefont{and}
  \bibinfo{author}{\bibfnamefont{A.~S.} \bibnamefont{Mikhailov}},
  \bibinfo{journal}{Phys. Rev. Lett.} \textbf{\bibinfo{volume}{93}},
  \bibinfo{pages}{254101} (\bibinfo{year}{2004}).

\bibitem[{\citenamefont{Restrepo et~al.}(2006)\citenamefont{Restrepo, Ott, and
  Hunt}}]{restr06}
\bibinfo{author}{\bibfnamefont{J.~G.} \bibnamefont{Restrepo}},
  \bibinfo{author}{\bibfnamefont{E.}~\bibnamefont{Ott}}, \bibnamefont{and}
  \bibinfo{author}{\bibfnamefont{B.~R.} \bibnamefont{Hunt}},
  \bibinfo{journal}{Physical Review Letters} \textbf{\bibinfo{volume}{96}},
  \bibinfo{pages}{128101} (\bibinfo{year}{2006}).

\bibitem[{\citenamefont{Restrepo et~al.}(2005)\citenamefont{Restrepo, Hunt, and
  Ott}}]{restr05}
\bibinfo{author}{\bibfnamefont{J.~G.} \bibnamefont{Restrepo}},
  \bibinfo{author}{\bibfnamefont{B.~R.} \bibnamefont{Hunt}}, \bibnamefont{and}
  \bibinfo{author}{\bibfnamefont{E.}~\bibnamefont{Ott}},
  \bibinfo{journal}{Physical Review E} \textbf{\bibinfo{volume}{71}},
  \bibinfo{pages}{036151} (\bibinfo{year}{2005}).

\bibitem[{\citenamefont{Grigoriev et~al.}(1997)\citenamefont{Grigoriev, Cross,
  and Schuster}}]{Pinn1}
\bibinfo{author}{\bibfnamefont{R.}~\bibnamefont{Grigoriev}},
  \bibinfo{author}{\bibfnamefont{M.}~\bibnamefont{Cross}}, \bibnamefont{and}
  \bibinfo{author}{\bibfnamefont{H.}~\bibnamefont{Schuster}},
  \bibinfo{journal}{Phys. Rev. Lett.} \textbf{\bibinfo{volume}{79}},
  \bibinfo{pages}{2795} (\bibinfo{year}{1997}).

\bibitem[{\citenamefont{Parekh et~al.}(1998)\citenamefont{Parekh,
  Parthasarathy, and Sinha}}]{Pinn2}
\bibinfo{author}{\bibfnamefont{N.}~\bibnamefont{Parekh}},
  \bibinfo{author}{\bibfnamefont{S.}~\bibnamefont{Parthasarathy}},
  \bibnamefont{and} \bibinfo{author}{\bibfnamefont{S.}~\bibnamefont{Sinha}},
  \bibinfo{journal}{Phys. Rev. Lett.} \textbf{\bibinfo{volume}{81}},
  \bibinfo{pages}{1401} (\bibinfo{year}{1998}).

\bibitem[{\citenamefont{Wang and Chen}(2002)}]{PinnA}
\bibinfo{author}{\bibfnamefont{X.}~\bibnamefont{Wang}} \bibnamefont{and}
  \bibinfo{author}{\bibfnamefont{G.}~\bibnamefont{Chen}},
  \bibinfo{journal}{Physica A} \textbf{\bibinfo{volume}{310}},
  \bibinfo{pages}{521} (\bibinfo{year}{2002}).

\bibitem[{\citenamefont{Li et~al.}(2004)\citenamefont{Li, Wang, and
  Chen}}]{PinnIEEE}
\bibinfo{author}{\bibfnamefont{X.}~\bibnamefont{Li}},
  \bibinfo{author}{\bibfnamefont{X.}~\bibnamefont{Wang}}, \bibnamefont{and}
  \bibinfo{author}{\bibfnamefont{G.}~\bibnamefont{Chen}},
  \bibinfo{journal}{IEEE Transa. on Circuits and Systems -I}
  \textbf{\bibinfo{volume}{51}}, \bibinfo{pages}{2074} (\bibinfo{year}{2004}).

\bibitem[{\citenamefont{Barabasi and Albert}(1999)}]{Ba:Al99}
\bibinfo{author}{\bibfnamefont{A.}~\bibnamefont{Barabasi}} \bibnamefont{and}
  \bibinfo{author}{\bibfnamefont{R.}~\bibnamefont{Albert}},
  \bibinfo{journal}{Science} \textbf{\bibinfo{volume}{286}},
  \bibinfo{pages}{509} (\bibinfo{year}{1999}).

\bibitem[{\citenamefont{Goh et~al.}(2001)\citenamefont{Goh, Kahng, and
  D.Kim}}]{korea}
\bibinfo{author}{\bibfnamefont{K.-I.} \bibnamefont{Goh}},
  \bibinfo{author}{\bibfnamefont{B.}~\bibnamefont{Kahng}}, \bibnamefont{and}
  \bibinfo{author}{\bibnamefont{D.Kim}}, \bibinfo{journal}{Phys. Rev. Lett.}
  \textbf{\bibinfo{volume}{87}} (\bibinfo{year}{2001}).

\bibitem[{\citenamefont{R.Pastor-Satorras and A.Vespignani}(2001)}]{Pa:Ve00a}
\bibinfo{author}{\bibnamefont{R.Pastor-Satorras}} \bibnamefont{and}
  \bibinfo{author}{\bibnamefont{A.Vespignani}}, \bibinfo{journal}{Phys. Rev.
  Lett.} \textbf{\bibinfo{volume}{86}} (\bibinfo{year}{2001}).

\bibitem[{\citenamefont{Girvan and Newman}(2002)}]{Ne:Gi02}
\bibinfo{author}{\bibfnamefont{M.}~\bibnamefont{Girvan}} \bibnamefont{and}
  \bibinfo{author}{\bibfnamefont{M.}~\bibnamefont{Newman}},
  \bibinfo{journal}{PNAS} \textbf{\bibinfo{volume}{99}}, \bibinfo{pages}{7821}
  (\bibinfo{year}{2002}).

\bibitem[{\citenamefont{Newman}(2003)}]{New03Mix}
\bibinfo{author}{\bibfnamefont{M.}~\bibnamefont{Newman}},
  \bibinfo{journal}{Phys. Rev. E} \textbf{\bibinfo{volume}{67}}
  (\bibinfo{year}{2003}).

\bibitem[{\citenamefont{Newman}(2002)}]{New02Ass}
\bibinfo{author}{\bibfnamefont{M.}~\bibnamefont{Newman}},
  \bibinfo{journal}{Phys. Rev. Lett.} \textbf{\bibinfo{volume}{89}}
  (\bibinfo{year}{2002}).

\bibitem[{\citenamefont{Sorrentino et~al.}(2006)\citenamefont{Sorrentino,
  di~Bernardo, Huerta-Cuellar, and Boccaletti}}]{So:di:Bo}
\bibinfo{author}{\bibfnamefont{F.}~\bibnamefont{Sorrentino}},
  \bibinfo{author}{\bibfnamefont{M.}~\bibnamefont{di~Bernardo}},
  \bibinfo{author}{\bibfnamefont{G.}~\bibnamefont{Huerta-Cuellar}},
  \bibnamefont{and}
  \bibinfo{author}{\bibfnamefont{S.}~\bibnamefont{Boccaletti}},
  \bibinfo{journal}{Physica D} \textbf{\bibinfo{volume}{224}},
  \bibinfo{pages}{123} (\bibinfo{year}{2006}).

\bibitem[{\citenamefont{R.Xulvi-Brunet and I.M.Sokolov}(2004)}]{reshuffling}
\bibinfo{author}{\bibnamefont{R.Xulvi-Brunet}} \bibnamefont{and}
  \bibinfo{author}{\bibnamefont{I.M.Sokolov}}, \bibinfo{journal}{Phys. Rev. E.}
  \textbf{\bibinfo{volume}{70}} (\bibinfo{year}{2004}).

\bibitem[{\citenamefont{Oh et~al.}(2005)\citenamefont{Oh, Rho, Hong, and
  Kahng}}]{Oh:Rho}
\bibinfo{author}{\bibfnamefont{E.}~\bibnamefont{Oh}},
  \bibinfo{author}{\bibfnamefont{K.}~\bibnamefont{Rho}},
  \bibinfo{author}{\bibfnamefont{H.}~\bibnamefont{Hong}}, \bibnamefont{and}
  \bibinfo{author}{\bibfnamefont{B.}~\bibnamefont{Kahng}},
  \bibinfo{journal}{Phys. Rev. E} \textbf{\bibinfo{volume}{72}},
  \bibinfo{pages}{047101} (\bibinfo{year}{2005}).

\bibitem[{\citenamefont{Arenas et~al.}(2006)\citenamefont{Arenas, Diaz-Guilera,
  and Perez-Vicente}}]{Ar:Di:PV}
\bibinfo{author}{\bibfnamefont{A.}~\bibnamefont{Arenas}},
  \bibinfo{author}{\bibfnamefont{A.}~\bibnamefont{Diaz-Guilera}},
  \bibnamefont{and} \bibinfo{author}{\bibfnamefont{C.~J.}
  \bibnamefont{Perez-Vicente}}, \bibinfo{journal}{Phys. Rev. Lett.}
  \textbf{\bibinfo{volume}{96}}, \bibinfo{pages}{114102}
  (\bibinfo{year}{2006}).

\bibitem[{\citenamefont{Park et~al.}(2006)\citenamefont{Park, Lai, and
  Gupte}}]{Pa:Lai}
\bibinfo{author}{\bibfnamefont{K.}~\bibnamefont{Park}},
  \bibinfo{author}{\bibfnamefont{Y.-C.} \bibnamefont{Lai}}, \bibnamefont{and}
  \bibinfo{author}{\bibfnamefont{S.}~\bibnamefont{Gupte}},
  \bibinfo{journal}{Chaos} \textbf{\bibinfo{volume}{16}},
  \bibinfo{pages}{015105} (\bibinfo{year}{2006}).

\bibitem[{\citenamefont{Nishikawa and Motter}(2006)}]{Ni:Mo06}
\bibinfo{author}{\bibfnamefont{T.}~\bibnamefont{Nishikawa}} \bibnamefont{and}
  \bibinfo{author}{\bibfnamefont{A.~E.} \bibnamefont{Motter}},
  \bibinfo{journal}{Phys. Rev. E.} \textbf{\bibinfo{volume}{73}},
  \bibinfo{pages}{065106} (\bibinfo{year}{2006}).

\bibitem[{\citenamefont{Newman and Girvan}(2004)}]{Ne:Gi04}
\bibinfo{author}{\bibfnamefont{M.}~\bibnamefont{Newman}} \bibnamefont{and}
  \bibinfo{author}{\bibfnamefont{M.}~\bibnamefont{Girvan}},
  \bibinfo{journal}{Phys. Rev. E} \textbf{\bibinfo{volume}{69}}
  (\bibinfo{year}{2004}).

\end{thebibliography}
\end{document}